\documentclass[11pt]{article}

\usepackage[utf8]{inputenc}

\usepackage[pdftex]{graphicx}

\usepackage[margin=1in]{geometry}

\usepackage{authblk} 

\usepackage{amsmath,amssymb,amsthm}

\newtheorem{theorem}{Theorem}
\newtheorem{corollary}[theorem]{Corollary}
\newtheorem{lemma}{Lemma}
\newtheorem{definition}{Definition}

\newcommand{\hide}[1]{}

\usepackage{subfig}
\usepackage{graphicx}

\newcommand{\Tn}{\ensuremath{\mathcal{T}_{n}}}
\newcommand{\Gn}{\ensuremath{\mathcal{G}_{n}}}

\newcommand{\tp}[1]{\ensuremath{U_{#1}}}
\newcommand{\bt}[1]{\ensuremath{L_{#1}}}

\newcommand{\N}{\ensuremath{\mathbb{N}}}

\newcommand{\subtr}{\ensuremath{\sqsubseteq}}
\newcommand{\trept}[2]{\ensuremath{\sigma(#1, #2)}}
\newcommand{\fiso}{\ensuremath{\simeq_0}}
\newcommand{\biso}{\ensuremath{\simeq_1}}

\newcommand{\desc}[3]{\ensuremath{\delta_{#1}(#2, #3)}}
\newcommand{\notdesc}[3]{\ensuremath{\bar{\delta}_{#1}(#2, #3)}}

\newcommand{\mmx}[2]{\ensuremath{#1 \rightarrow #2}}
\newcommand{\notmmx}[2]{\ensuremath{#1 \not\rightarrow #2}}

\newcommand{\mcp}[2]{\ensuremath{\pi(#1,#2)}}
\newcommand{\glb}[2]{\ensuremath{\gamma(#1,#2)}}
\newcommand{\lub}[2]{\ensuremath{\lambda(#1,#2)}}

\newcommand{\NCn}{\ensuremath{{\rm NC}(n)}}
\newcommand{\nc}[1]{\ensuremath{\mathcal{N}_{#1}}}

\newcommand{\ind}[1]{\ensuremath{\alpha(#1)}}
\newcommand{\indeven}[1]{\ensuremath{\alpha_2(#1)}}
\newcommand{\indodd}[1]{\ensuremath{\alpha_1(#1)}}
\newcommand{\even}[2]{\ensuremath{\epsilon(#1, #2)}}

\newcommand{\polat}{\ensuremath{(O_n, \cap)}}

\newcommand{\abs}[1]{\ensuremath{\lvert #1 \rvert}}

\begin{document}

\title{\vspace*{-2ex}On a barrier height problem for RNA branching}
 
\author[1]{Christine Heitsch}
\author[2]{Chi N.Y.\ Huynh}
\author[3]{Greg Johnston}
\affil[1]{School of Mathematics, Georgia Institute of Technology}
\affil[2]{Analysis Group, Inc.}
\affil[3]{Rubrik, Inc.}

\renewcommand\Authands{ and }

\date{MSC 92D20 (05C05, 05A18) }
 
\maketitle

The branching of an RNA molecule is an important 
structural characteristic yet difficult to predict correctly, 
especially for longer sequences.
Using plane trees as a combinatorial model for RNA folding,
we consider the thermodynamic cost, known as the barrier height,
of transitioning between branching configurations.
Using branching skew as a coarse energy approximation, we characterize
various types of paths in the discrete configuration landscape.
In particular,
we give sufficient conditions for a path to have both minimal length 
and minimal branching skew.
The proofs offer some biological insights, notably the potential 
importance of both hairpin stability and domain architecture
to higher resolution RNA barrier height analyses.

\section{Introduction}

An RNA sequence is said to fold into a secondary structure via the
formation of (noncrossing, canonical) base pairings.
There are many possible secondary structures for a given
sequence, but the most biologically relevant typically
have a low free energy approximation under the nearest neighbor
thermodynamic model (NNTM).
The barrier height problem~\cite{morgan-higgs-98} then 
considers the thermodynamic cost of transitioning between low-energy
configurations.
Progress has typically 
focused on steps consisting of adding/removing a base pair, 
c.f.~\cite{dotu-etal-10, li-zhang-12, takizawa-etal-20} and 
related work discussed therein. 
Here, we take a complementary approach, focusing on larger 
structural rearrangements by using plane trees as a 
combinatorial model of RNA branching configurations.

A plane tree is a rooted tree whose subtrees are linearly 
ordered~\cite{stanley-99}.
Also know as ordered or linear trees, they are one of the 
many combinatorial families enumerated by the Catalan numbers.
Depending on the question of interest,
there are different ways\footnote{See~\cite{insights_chapt}
for an overview of the combinatorics of RNA secondary structures
and more comprehensive references.}
of associating RNA secondary structures
with trees in general, e.g.~\cite{gan-pasquali-schlick-03},
and plane trees in particular, e.g.~\cite{schmitt-waterman-94}.
As done in other branching  
analyses~\cite{ldpmath, ldpbio, dna8, parametric},
we take a low-resolution approach,
associating helices to edges and loops to vertices
with the external loop as the distinguished root vertex.

A plane tree is thus an abstract representation of 
an arbitrary RNA secondary structure.
By focusing on the overall arrangement of edges/helices
and vertices/loops, 
mathematical results have provided insight into  
the challenge of designing RNA sequences with 
a particular branching structure~\cite{dna8}, 
configurations which minimize loop energy 
costs~\cite{ldpmath, ldpbio},
and a parametric analysis of the branching 
entropy approximation~\cite{parametric}.
This work has lead both to better understanding of 
RNA prediction accuracy~\cite{regions, robust, bnb} 
as well as some new combinatorics~\cite{fpsac}.

Here, we extend this theoretical branching analysis to consider 
folding pathways between plane trees.
We move from one tree to another under a ``pairing exchange'' 
operation inspired by the challenge of encoding a particular
branched structure in a sequence~\cite{insights_chapt, dna8}.
Under a coarse approximation to the thermodynamics (branching skew),
combinatorial analysis of different types of transition paths is possible
in this model of RNA folding.
The proofs offer some biological insights.

First, there is a direct path between any two trees, i.e.\ one where 
each step increases the number of edges from the final tree by at
least one.
The edges incident on a leaf are the crucial first steps in 
such a path, and indeed the stability of RNA hairpins is a 
critical component of biological function~\cite{bevilacqua-blose-08}
and modeling accuracy~\cite{sulc-20}.
Hence, a suitable model for hairpin rearrangements~\cite{xu-chen-12} may be 
an important component of a higher resolution barrier height analysis.

Second, the branching skew of a direct path is provably bounded
when the edges of the two trees decompose into consistent blocks
that can rearrange from initial to final configurations independent
of each other.
This suggests that modeling the domain architecture of RNA 
secondary structures,
which emerges in the folding of longer 
sequences~\cite{huston-etal-21,petrov-etal-13,quinn-etal-14}, 
may be critical to the analysis of optimal folding pathways.

\section{Pairing exchanges and branching skew}\label{sec:prelim}

Let \Tn\ denote the set of plane trees with $n$ edges.
Then $|\Tn| = \frac{1}{n+1}\binom{2n}{n}$, the $n$-th Catalan number.
Motivated by RNA secondary structures, we consider $T \in \Tn$ to be a set 
of paired half-edges.
For $i, j \in \N$, let $[i,j] = \{k \in \N \mid i \leq k \leq j\}$.
Label the boundary of $T$ counter-clockwise from the root
with $[1,2n]$ in increasing order.
Let $(i,j)$ denote the edge in $T$ which has $i$ as the label on its
left side and $j$ on the right for $1 \leq i < j \leq 2n$.

\begin{lemma}
A set $I = \{(i,j) \mid 1 \leq i < j \leq 2n\}$ is a plane tree
when each index appears in exactly one ordered pair and there do
not exist $(i,j), (i',j') \in I$ with $i < i' < j < j'$.
\end{lemma}

\begin{proof}
Consider $(1, k) \in I$. 
If $k$ is odd, then either there exists an $(i,j) \in I$ with $1 < i < k < j$
or an index in $[2,k-1]$ that is unpaired or in more than one pairing.
Since $k$ must be even, induct on the pairings with indices in 
$[2,k-1]$ and $[k+1,2n]$.
\end{proof}

In other words, there is a simple bijection between noncrossing perfect
matchings on $2n$ endpoints and plane trees with $n$ edges.
Previous work~\cite{fpsac} considered the comparable operation on 
matchings to the pairing exchange defined below with the goal of 
better understanding meanders 
(interpreted as pairs of noncrossing perfect matchings which form
a single closed loop). 

Here we consider plane trees as a low-resolution model
of RNA secondary structures,
and analyze (very approximately) the thermodynamic cost of 
moving around this branching configuration landscape.
Inspired by the challenge of minimizing alternative lower-energy 
configurations when designing RNA secondary structures
(c.f.\ Fig.~1 in \cite{dna8}),
we transition from one tree to the next by breaking apart and ``repairing'' 
two edges. 

We start by applying to edges in $T$ 
the common familial terminology for vertices in rooted trees, 
i.e.\ parent/child, siblings, ancestor/descendent, etc.
Additionally, an edge incident on the root vertex is called an orphan.
Two edges in $T$ are \emph{unobstructed} if they are
incident on the same vertex, in which case they are either parent/child 
or siblings.

We define a \emph{pairing exchange} on unobstructed edges 
$E = \{(i,j), (i',j')\} \subseteq T$ as 
\[
\mu_E(T) = 
(T \setminus E) \cup 
\left \{
\begin{array}{cc} 
(i, i') \mbox{ and } (j', j) & \mbox{ if } i < i' < j' < j \\
(i, j') \mbox{ and } (j,i')) & \mbox{ if } i < j < i' < j' 
\end{array} 
\right \} 
\]
and claim that converting a parent/child into siblings, 
or vice versa, introduces no crossings.

\begin{figure}
\centering
\includegraphics[width = .5\textwidth]{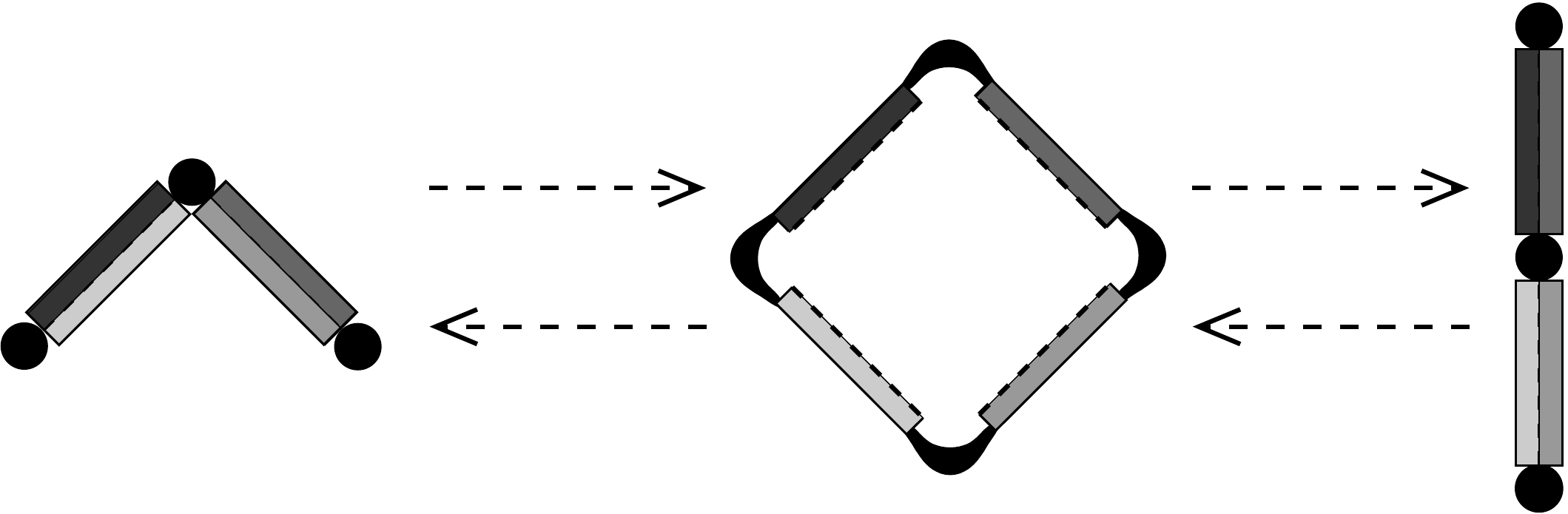}
\caption{A pairing exchange on unobstructed edges with 
indices $1 \leq a < b < c < d \leq 2n$
converts siblings $(a,b)$, $(c,d)$ into parent/child 
$(a,d)$, $(b,c)$, and vice versa.
Note how the incident vertices split and merge.
However, edges in the four subtrees with indices exclusively in 
$A = [1,a-1] \cup [d+1,2n]$,  
$B = [a+1,b-1]$, $C = [b+1,c-1]$, or $D = [c+1,d-1]$ 
are unaltered.
\label{fig:pairingexch}}
\end{figure}

\begin{lemma}\label{lem:exdef}
The pairing exchange operation is well-defined.
\end{lemma}

\begin{proof}
Let $1 \leq a < b < c < d \leq 2n$ be the indices of two edges in $T$.
Let $A = [1,a-1] \cup [d+1,2n]$, 
$B = [a+1,b-1]$, $C = [b+1,c-1]$, and $D = [c+1,d-1]$.
Observe that if the edges are $(a,d)$ and $(b,c)$, then all other edges
must have indices in either $A$ or in $B \cup D$ or in $C$ exclusively.
However, if $(a,d)$ and $(b,c)$ are parent and child, then there
cannot be an edge $(k,l)$ with $k \in B$ and $l \in D$.
A similar argument holds if $(a,b)$ and $(c,d)$ are siblings.
\end{proof}

As illustrated in Figure~\ref{fig:pairingexch},
pairing exchanges are reversible operations.  
Let \Gn\ be the (undirected) graph with vertex set \Tn\ and edges 
which connect two plane trees that differ by a single pairing exchange.

Before proving that \Gn\ is connected,
we distinguish two trees which have the maximum degree of 
$\binom{n}{2}$ in \Gn.
Let $\tp{n} = \{(2i-1,2i) \mid 1 \leq i \leq n\}$
and $\bt{n} = \{(1,2n)\} \cup \{(2i,2i+1) \mid 1 \leq i \leq n-1\}$.
Then $\tp{1} = \bt{1}$, and $\mathcal{G}_1$ consists of a single vertex.
For $n \geq 2$,
$\tp{n}$ and $\bt{n}$ differ in the choice of root; 
both are ``star'' trees with $n$ unobstructed edges.

\begin{figure}
\centering
\includegraphics[width = .5\textwidth]{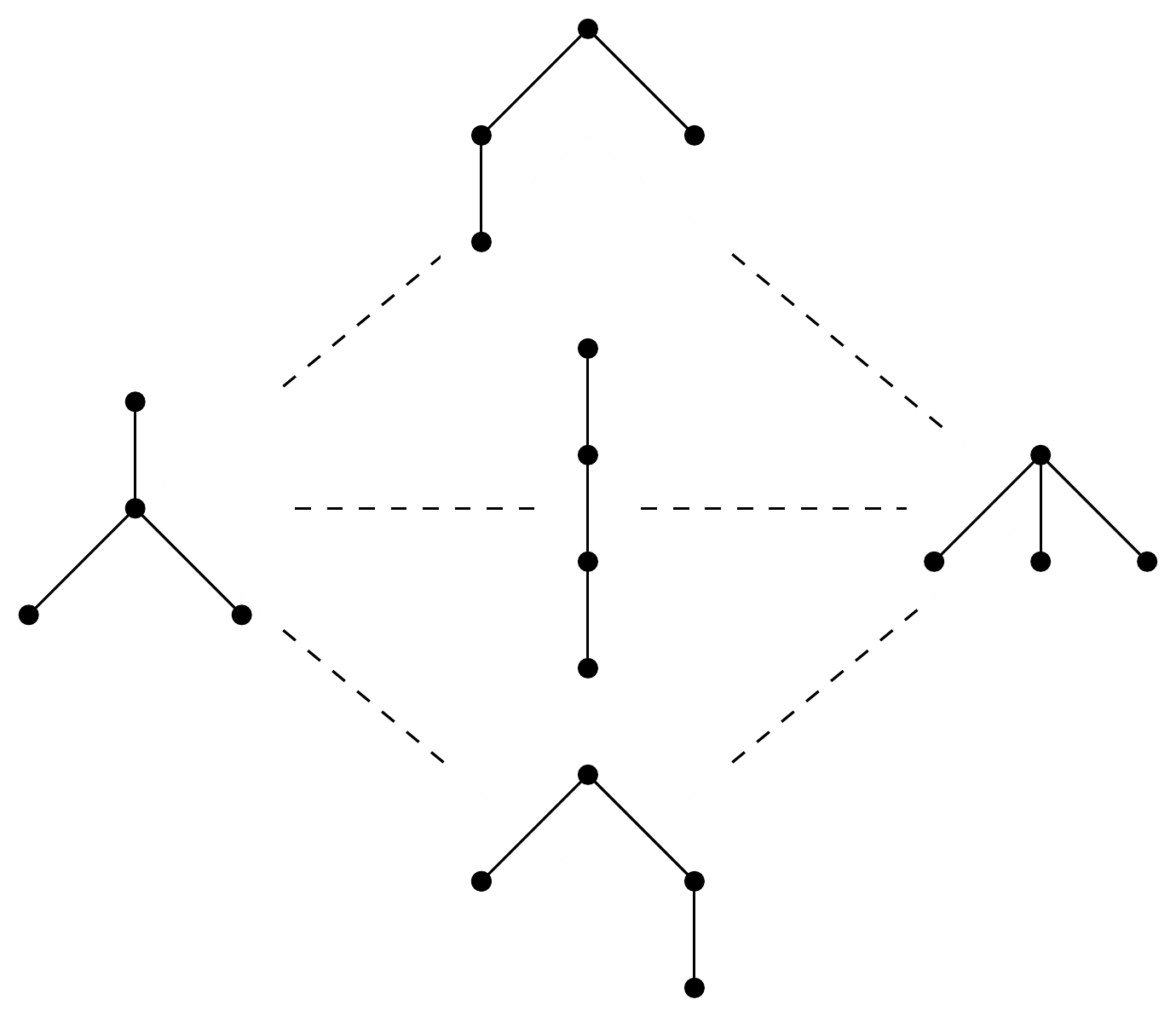}
\caption{The graph $\mathcal{G}_3$ with $\bt{3}$ on left, 
$\tp{3}$ on right, and the three plane trees $T \in \mathcal{T}_3$
with $c(T) = 2$ in the middle.
Dashed lines are pairing exchanges.
The number of odd edges increases by 1 moving left to right.
\label{fig:g3top}}
\end{figure}

\begin{lemma}\label{lem:connect}
The graph \Gn\ is connected.
\end{lemma}

\begin{proof}
We claim there is a path in \Gn\ from $T$ to $\tp{n}$.
If $(1,2) \in T$, then inductively the subtree with  
indices in $[3,2n]$ is connected by pairing exchanges to $\tp{n-1}$.
Else the edges $(1,k), (2,l) \in T$ are unobstructed, and there is an
edge in \Gn\ between $T$ and 
$[T \setminus \{(1,k), (2,l)\}] \cup \{(1,2), (k,l)\}$.
\end{proof}

Dually, $T$ is connected to $\bt{n}$ by
first considering the edge $(1,2n)$, and then successive $(2i, 2i+1)$.
Thus, any two trees are connected by a path through $\tp{n}$ or 
one through $\bt{n}$.
Unless a star tree is one of the endpoints, the path is indirect
since it effectively erases most pairing information in the first
tree before replacing it with second.
Moreover, although mathematically simple, 
these are the highest possible barrier paths in terms of branching
thermodynamics.

Previous results~\cite{ldpbio, insights_chapt, parametric} demonstrated
that branching is locally favorable but globally balanced by 
increasing the number of leaves ---
since hairpins are the most energetically expensive type of 
loop structures~\cite{turner-mathews-10}.
Hence, a low barrier path in \Gn\ passes through trees with 
a low degree of branching.
Since tracking changes in branching degree
under pairing exchanges is complicated, 
we instead consider ``branching skew.''

We start by defining the parity of an edge.
Since $(i,j) \in T$ has $\frac{j-i-1}{2}$ descendents,
exactly one of $i$ and $j$ is odd.
Call the edge \emph{odd} if $i$ is, and \emph{even} otherwise.
Let $c(T)$ denote the number of odd edges in $T$.
Then $1 \leq c(T) \leq n$,  
with $c(T) = 1$ exactly when $T = \bt{n}$ and $n$ only if $\tp{n}$.
Moreover, a pairing exchange alters the number of odd edges by exactly one.   
An example is seen in Figure~\ref{fig:g3top}.

Note that edge parity along a path in $T$ from the root vertex 
to a leaf must alternate, and all orphan edges are odd.
Hence, if $c(T) = k$, then the maximum possible vertex degree
in $T$ is either $k$ or $n - k + 1$.
A tree which achieves this is $k$ orphans
with one having $n - k$ even children.
Thus, $c(T)$ is well-behaved under pairing exchanges, and
yields an upper bound, seldom tight, on vertex degree.

More precisely, we are interested in minimizing the maximum possible
vertex degree over paths in \Gn.
Define the \emph{skew} of $T$ to be 
$\abs{c(T) - \frac{n+1}{2}}$.
This is maximal at both $\tp{n}$ and $\bt{n}$, and decreases to 
a minimum of $0$ or $1/2$ when 
the number of odd and even edges are most evenly balanced.
The skew of a path $T = T_0$, $T_1$, \ldots, $T_k = T'$
is $\max_{0 \leq m \leq k} \abs{c(T_m) - \frac{n+1}{2}}$.

Call a pairing exchange a \emph{forward} move if $c(T)$ increases,
and \emph{backward} otherwise.
In Section~\ref{sec:forward}, we characterize when there is a 
path from $T$ to $T'$ consisting only of forward moves.
Call this a forward path, and the reverse a backward one.
Such a path has the least increase in skew possible given 
the start and end.

Lemma~\ref{lem:connect} showed that,
even if there is not a forward
path from $T$ to $T'$, they are still connected
by a pair of forward paths through $\tp{n}$.
Dually, backward through $\bt{n}$.
More generally, we call a path from $T$ to $T'$ 
a \emph{forward V-path} (respectively \emph{backward}) 
if there exists $S \in \Tn$ with $S \neq T, T'$
such that there is a forward (resp. backward) path from $T$ to $S$ 
and also from $T'$ to $S$.
In Section~\ref{sec:vee}, we characterize the minimal skew V-paths,
and make explicit in Section~\ref{sec:order} 
the connection with the well-studied lattice of noncrossing partitions.

In Section~\ref{sec:bounded}, we show that when forward and backward 
moves are interspersed, it is possible to have paths in \Gn\ whose skew
exceeds the start and end by at most 1.
We conclude in Section~\ref{sec:geodesic} by characterizing shortest
paths, and proving that their skew is similarly bounded
under certain conditions.

\section{Introducing tree partitions}\label{sec:partition}

Since a plane tree $T$ is specified as a collection of paired half-edges 
$\{(i,j) \mid 1 \leq i < j \leq 2n\}$, we distinguish when a 
subset $S \subseteq T$ is a subtree, denoted $S \subtr T$.
When $S$ is connected,
by a generalization of Lemma~\ref{lem:exdef}, 
an edge in $T \setminus S$ has either both or neither indices 
in an interval between the ordered indices of $S$.
Hence, pairing exchanges on $S$ and its
subsequent images are independent of all other edges in $T$.

Because each edge has exactly one odd index,
subsets of $T \in \Tn$ are in bijection with subsets of 
$O_n = \{1, 3, \ldots, 2n - 1\}$.
For $P \subseteq O_n$, let 
$\trept{T}{P} = \{ (i,j) \in T \mid i \in P \mbox{ or } j \in P\}$.
Let $\mathcal{P}$ be a (set) partition of $O_n$. 
We distinguish when the parts of $\mathcal{P}$ decompose $T$ into subtrees.

\begin{definition}
Say $\mathcal{P}$ \emph{splits} $T$ 
if $\trept{T}{P} \subtr T$ for every $P \in \mathcal{P}$.
\end{definition}

\noindent
The trees $\tp{n}$ and $\bt{n}$ are split by any $\mathcal{P}$,
since all edges are incident on a common vertex.
However, suppose there exists $P \in \mathcal{P}$ and 
odd integers $i < j < k$ (circularly ordered) such that
$i, k \in P$ and $j \notin P$.
Let $T = \mu_E(\tp{n})$ for $E = \{(j,j+1), (k,k+1)\}$.
Then $(j, k+1)$ obstructs $(j+1, k)$ from $(i, i+1)$ in $T$, 
and so $\mathcal{P}$ does not split $T$.
Hence, there are exactly two partitions which split every $T \in \Tn$:
$\{O_n\}$ and $\{ \{2i - 1\} \mid 1 \leq i \leq n\}$.
The latter will be referred to as the singleton partition, 
and the former as the trivial one.

To characterize paths in \Gn\ from $T$ to $T'$, 
we consider partitions of $O_n$ which split both trees.
However, we must insure that the even indices also 
partition in the same way.
For $S \subseteq T$, let $\ind{S} = \bigcup_{(i,j) \in S} \{i, j\}$
be the collection of indices.
Denote the odd ones by $\indodd{S}$, respectively 
$\indeven{S}$ for the even.

\begin{definition}
The subsets $S \subseteq T$, $S' \subseteq T'$ are \emph{aligned} if 
$\ind{S} = \ind{S'}$.
The alignment is \emph{simple} if no proper subsets of $S$ and $S'$
are also aligned.
\end{definition}

\noindent
The simply aligned subsets correspond to connected components in 
the graph with vertices in $[1,2n]$ and edges in $T \cup T'$.
It is known~\cite{fpsac} that
a pairing exchange either splits a connected component into two 
or joins two disjoint ones.
Hence, the number of simply aligned subsets changes by exactly 1
across each edge $\{T, T'\} \in \Gn$.

Observe that if $T$ and $T'$ decompose into $k$ pairs of aligned subtrees
$S_m \subtr T$, $S'_m \subtr T'$, then there is a path in \Gn\ from 
$T = T_0$ to $T' = T_k$ through $T_{m} = (T_{m-1} \setminus S_m) \cup S'_m$.
In other words, pairing exchanges on distinct aligned subtrees are 
independent.

Since $\trept{T}{P}$ and $\trept{T'}{P}$ have the same odd indices by 
definition, let $\even{T}{P} = \indeven{\trept{T}{P}}$. 
Then the induced subtrees are aligned exactly when 
$\even{T}{P} = \even{T'}{P}$.

\begin{definition}
Let $\mathcal{P}$ be a partition of $O_n$ and $\mathcal{S} \subseteq \Tn$.
Suppose $\mathcal{P}$ splits every $T \in \mathcal{S}$.
Suppose further that $\even{T}{P} = \even{T'}{P}$ for every 
$T, T' \in \mathcal{S}$ and $P \in \mathcal{P}$.
Then $\mathcal{P}$ is a \emph{tree partition of $\mathcal{S}$}.
\end{definition}

\noindent
While the trivial partition meets the alignment criteria for any 
$T$ and $T'$, the singleton one fails unless $T = T'$.

Recall that set partitions form a lattice, partially ordered under 
refinement.
Here we take the singleton partition of $O_n$ as the minimum element
since it induces subtrees with the fewest number of edges.
The trivial partition, which is a tree partition for any  
$\mathcal{S} \subseteq \Tn$, is then the maximum.
This lattice will be denoted \polat. 

\begin{lemma}
For $\mathcal{S} \subseteq \Tn$, 
there is a unique tree partition of $\mathcal{S}$,
denoted $\pi(\mathcal{S})$,
minimal in \polat.
\end{lemma}

\begin{proof}
Suppose $\mathcal{Q} \neq \mathcal{Q'}$ are both minimal
and let $\mathcal{P}$ be their greatest lower bound under refinement.
Let $P \in \mathcal{P}$.
Then $P = Q \cap Q'$ for some $Q \in \mathcal{Q}$, $Q' \in \mathcal{Q'}$.

Let $T \in \mathcal{S}$, and 
suppose $(k,l) \in T$ lies on the path
between $(i,j), (i',j') \in \trept{T}{P}$.
Since $\trept{T}{X} \subtr T$ for $X = Q, Q'$,
then $(k,l) \in \trept{T}{P}$.
Since the edges in $T$ with odd endpoints in $P$ are connected,
$\mathcal{P}$ splits $T$.

Let $i \in \even{T}{P}$.  
Let $T' \in \mathcal{S}$, and suppose $i$ pairs with $j$ in $T'$.
Since $\even{T}{Q} = \even{T'}{Q}$, then $j \in Q$.  
Likewise for $Q'$.  
Hence $j \in P$, and $i \in \even{T'}{P}$.
Since the induced subtrees are aligned,
$\mathcal{P}$ is a tree partition of $\mathcal{S}$.
Contradiction.
\end{proof}

When $\mathcal{S} = \{T, T'\}$, write $\mcp{T}{T'}$.
To produce $\mcp{T}{T'}$, 
we can start with the simply aligned subsets.  
For example, consider 
$T = \{(1,8), (2,7), (3,6), (4,5)\}$ and 
$T' = \{(1,4), (2,3), (5, 8), (6,7)\}$.
The aligned subsets have $\ind{S} = \{1, 4, 5, 8\}$,
$\ind{S'} = \{2, 3, 6, 7\}$ and induce the partition of $O_n$ with 
parts $\indodd{S}$, $\indodd{S'}$.
However, $\trept{T}{\{1,5\}} \not\subtr T$ 
and $\trept{T'}{\{3,7\}} \not\subtr T'$.
To obtain a partition which also splits these trees, it suffices to take
the union of $\indodd{S}$ and $\indodd{S'}$.

More generally, let $\{S_i\}$ be the simply aligned subsets 
for $\mathcal{S} \subseteq \Tn$, 
i.e.\ connected components in the graph on $[1,2n]$
with all edges from $T \in \mathcal{S}$.
Then $\mathcal{P} = \{\indodd{S_i} \}$ 
satisfies the tree partition alignment condition by definition.
Moreover, any enlargement of $\mathcal{P}$ in \polat\ is still aligned.
If $\trept{T}{P}$ is not connected for some $T \in \mathcal{S}$,
$P \in \mathcal{P}$, then there is an edge $(k,l) \notin \trept{T}{P}$
on the path in $T$ between some $(i,j), (i',j') \in \trept{T}{P}$.
But this can be addressed by enlarging $P$ to include 
$\indodd{S_i}$ where $(k,l) \in \trept{T}{\indodd{S_i}}$.
Inductively, 
$\pi(\mathcal{S})$ is the unique least enlargement of $\mathcal{P}$
where the induced subtrees are all connected.
Note also that $\pi(S)$ is an enlargement of $\pi(S')$ for every 
$S' \subset S$.

\section{Characterizing forward paths}\label{sec:forward}

We consider when $T$ is connected to $T'$ by a sequence of forward moves,
i.e.\ pairing exchanges which increase $c(T)$ by 1.
As proved in Lemma~\ref{lem:connect}, there is a forward path from 
$T$ to $\tp{n}$,
and dually one backward down to $\bt{n}$.
Since pairing exchange alters $c(T)$ by exactly 1, 
with $c(\bt{n}) = 1$ and $c(\tp{n}) = n$,
then the former has length $n - c(T)$, and the latter $c(T) - 1$.

Call two trees, like $\tp{n}$ and $\bt{n}$,
\emph{complementary} if there is only one simply aligned subset.
Such trees will be considered more generally in 
Section~\ref{sec:geodesic}.
Now we show there is a forward path from $T$ to $T'$
exactly when there is a tree partition 
which splits them into pairs of complementary ``star'' subtrees.

For $S \subseteq T$, let $c(S)$ denote the number of odd edges.
If $S$ is connected, then
$S \subtr T$ is isomorphic to $S' \in \mathcal{T}_{|S|}$
under an order-preserving bijection on its indices.
We distinguish whether edge parity is preserved or reversed, 
denoted $S \fiso S'$ or $S \biso S'$ respectively.
When preserved, $c(S) = c(S')$.
If reversed, $c(S) = |S| - c(S')$.

\begin{definition} 
The tree $T$ has a \emph{minmax decomposition} with $T'$,
denoted $\mmx{T}{T'}$, if there exists a
tree partition $\mathcal{P}$ such that,
for every $P \in \mathcal{P}$ with $|P| = p$,
$\trept{T}{P} \fiso \bt{p}$, $\trept{T'}{P} \fiso \tp{p}$ or 
$\trept{T}{P} \biso \tp{p}$, $\trept{T'}{P} \biso \bt{p}$.
\end{definition}

In other words, if the odd (and even) indices of $T$ and $T'$ partition
so that the induced subtrees are isomorphic to the star tree, with opposite
choices of root determined by the edge parity, 
then they have a minmax decomposition.
Note that $\mmx{T}{T}$ under the singleton
partition and $\tp{1} = \bt{1} = \{(1,2)\}$.
Dually, $\mmx{\bt{n}}{\tp{n}}$ under the trivial partition.
If $\mmx{T}{T'}$,
then the induced subtree of 
$T'$ has $p-1$ more odd edges than the one in $T$. 
Hence, $c(T) \leq c(T')$.

Call $(i,j) \in T \cap T'$ a \emph{common} edge.
Equivalently, $\{(i,j)\}$ is a simply aligned subset,
or the induced subtree for a singleton part of $\mcp{T}{T'}$.

\begin{theorem}\label{th:minmax}
There is a forward path from $T$ to $T'$ in \Gn\ if and only if $\mmx{T}{T'}$.
\end{theorem}

\begin{proof}

Let $T = T_0$, $T_1$, \ldots, $T_{k-1}$, $T_{k} = T'$ be a forward
path in \Gn.
At each step, either an odd parent/even child are
converted into two odd siblings, or two even siblings are
changed into an even parent/odd child.

Let $\mathcal{P}_1$ be the partition which is all singletons,
except for a doubleton $P$ which consists of the odd indices 
involved in the pairing exchange on $T$.
If the exchange started with parent/child,
then $\trept{T}{P} \fiso \bt{2}$ and $\trept{T_1}{P} \fiso \tp{2}$.
Otherwise, 
$\trept{T}{P} \biso \tp{2}$ and $\trept{T_1}{P} \biso \bt{2}$.

Assume after $m$ steps that $T$ has a minmax decomposition with $T_m$
for tree partition $\mathcal{P}_m$.
Then an induced subtree in $T_m$ 
contains either no even edges or exactly one even parent.
Hence the next forward pairing exchange necessarily involves 
edges associated with distinct parts.
Let $T_{m+1} = \mu_{E}(T_m)$ for $E = \{(i,j), (i',j')\}$.
Then 
$(i,j) \in \trept{T_m}{P}$ and $(i',j') \in \trept{T_m}{P'}$  
for $P, P' \in \mathcal{P}_m$ with $P \neq P'$, $|P| = p$ and $|P'| = p'$.

Suppose $(i,j), (i',j') \in T_m$ are even siblings 
with $i < j < i' < j'$.
Then $j \in P$ and $(i,j)$ is the parent in $\trept{T_m}{P} \biso \bt{p}$, 
and similarly for $(i',j')$ and $P'$.
After the pairing exchange, we have
$T_{m+1} = (T_m \setminus \{(i,j), (i',j')\}) \cup \{(i,j'), (j,i')\}$.
The even edge $(i,j')$ is the parent of $(j,i')$ as well as 
of all the odd children of $(i,j)$ in $\trept{T_m}{P}$
and of $(i',j')$ in $\trept{T_m}{P'}$. 
Hence $\trept{T_{m+1}}{P \cup P'}$ is a subtree of $T_{m+1}$ and
by construction $\biso \bt{p + p'}$.

Consider the partition 
$\mathcal{P}_{m+1} = (\mathcal{P}_m \setminus \{P, P'\}) \cup \{P \cup P'\}$.
If the even siblings comprising $\trept{T}{P}$ and $\trept{T}{P'}$ 
have the same parent, then $\trept{T}{P \cup P'} \subtr T$.
By construction the subtree is aligned with $\trept{T_{m+1}}{P \cup P'}$ 
and $\biso U_{p + p'}$.

Otherwise, let $(k,l) \in T$ be the odd parent of $\trept{T}{P} \biso \tp{p}$.
By the alignment criteria, $P \cup \even{T}{P} \subseteq [i,j]$
since $(i,j) \in \trept{T_m}{P}$ is the even parent. 
Hence, $1 \leq k < i < j < l \leq 2n$.
Without loss of generality, $(k,l)$ is not an ancestor of $\trept{T}{P'}$.
But then, since $i < i'$ by assumption,
$l \in [j+1, i'-1]$.

Let $Q \in \mathcal{P}_m$ such that 
$(k,l) \in \trept{T}{Q}$.
By induction, an induced subtree of $T$ 
has either zero or one odd edge.
Hence, $\trept{T}{Q} \fiso \bt{|Q|}$ and so $\trept{T_m}{Q} \fiso \tp{|Q|}$.
Let $K = [k, i-1]$ and $L = [j+1, l]$.
Suppose there is $(k',l') \in \trept{T_m}{Q}$ with $k' \in K$, $l' \in L$.
But this obstructs $(i,j)$ from $(i',j')$.
Hence there are an even number of indices from $\trept{T_m}{Q}$ in $K$ and 
in $L$.
But then by counting there is a child $(k',l') \in \trept{T}{Q}$ with 
$k' \in K$ and $l' \in L$, contradicting the choice of $(k,l)$.
Thus, if the pairing exchange on $T_m$ began with two even siblings, 
then $\mmx{T}{T_{m+1}}$.

Suppose instead $(i,j)$ is the odd parent of $(i',j')$ in $T_m$. 
Then $i < i' < j' < j$ and 
$T_{m+1} = (T_m \setminus \{(i,j), (i',j')\}) \cup \{(i,i'), (j',j)\}$.
Before the pairing exchange, 
although $\trept{T_m}{P'} \biso \bt{p'}$ as before,
$\trept{T_m}{P}$ may be either $\fiso \tp{p}$ or $\biso \bt{p}$. 
In either case, the new edges in $T_{m+1}$ are odd siblings, 
along with the former odd siblings of $(i,j)$ and odd children of $(i',j')$.
Hence, $\trept{T_{m+1}}{P \cup P'} \biso \bt{p + p'}$ or 
$\fiso \tp{p + p'}$ according to whether the
even parent of $(i,j)$ is in $\trept{T_m}{P}$ or not.

If the edges of $\trept{T}{P \cup P'}$ are not connected in $T$,
then by the same type of argument as above, we arrive at a contradiction.
Since $\trept{T}{P \cup P'} \subtr T$, then 
either $\biso \tp{p + p'}$ or $\fiso \bt{p + p'}$ respectively. 
Since all other parts of the partition were unchanged, $\mathcal{P}_{m+1}$
is a tree partition yielding a minmax decomposition for $T$ 
with $T_{m+1}$.

Conversely, suppose $\mmx{T}{T'}$ with tree partition $\mathcal{P}$.
Let $S = \trept{T}{P}$ and $S' = \trept{T}{P'}$
for $P \in \mathcal{P}$ with $|P| = p \geq 2$. 
Suppose $S \fiso \bt{p}$ and $S' \fiso \tp{p}$.
There is a forward path of length $p-1$ 
from $\bt{p}$ to $\tp{p}$ in $\mathcal{G}_p$.
Operating on the corresponding edges in $S$, while 
keeping $T \setminus S$ fixed, there is a forward
path from $T$ to $T'' = (T \setminus S) \cup S'$ in \Gn.
Dually, the backward path in $\mathcal{G}_p$ becomes a forward one
when $S \biso \tp{p}$ and $S' \biso \bt{p}$.
Then $T''$ has $p$ more common edges with $T'$ than $T$ does.
Inductively, the other pairs of induced subtrees are unchanged.
Hence $\mmx{T''}{T}$ for the tree partition $\mathcal{Q} = 
(\mathcal{P} \setminus \{P\}) \cup \{\{q\} \mid q \in P\}$.
\end{proof}

Suppose $\mmx{T}{T'}$. 
Then the forward path's branching skew is
$\max \{ \abs{c(T) - \frac{n+1}{2}}, \abs{c(T') - \frac{n+1}{2}} \}$
depending on whether $c(T) - 1 < n - c(T')$.
Hence, it has the least possible barrier height given 
the start and end points.
In this case, by construction, there is a bijection between 
parts of $\mcp{T}{T'}$ and simply aligned subsets.
So the path's length is $\sum |P| - 1 = n - k$,
when there are $k$ parts $P \in \mcp{T}{T'}$.
This is the shortest possible, and is generalized to geodesics between
all trees in Section~\ref{sec:geodesic}.
However, bounding the branching skew when $\notmmx{T}{T'}$ is more
challenging, and we consider several different types of paths.

\section{Characterizing minimal skew V-paths}\label{sec:vee}

Even if $\notmmx{T}{T'}$, they are still connected by  
a forward V-path of length $2n - c(T) - c(T')$ through $\tp{n}$, 
respectively backward of length $c(T) + c(T') - 2$ through $\bt{n}$.
These paths have the maximum possible skew 
of $\frac{n-1}{2}$,
and hence represent the highest barrier in branching thermodynamics.
However, this can be reduced in many cases
by restricting the rearrangements to suitable subtrees.

We beging by introducing some additional notation and terminology.
Let $\mathcal{P}$ be a tree partition of $\mathcal{S} \subseteq \Tn$.
For $P \in \mathcal{P}$, let $\min{P}$ be the least index in 
$P \cup \even{T}{P}$ and $\max{P}$ the greatest.
By the alignment criteria, these are well-defined.
Note they have opposite parity.
Call $P$ \emph{odd} if $\min{P}$ is, and \emph{even} otherwise.
Let $(i,j), (i', j') \in T$.
Call $(i',j')$ the \emph{first} child of $(i,j)$ if $i' = i+1$,
respectively \emph{last} if $j' + 1 = j$.
Say $(i', j')$ is the \emph{next} sibling of $(i,j)$ if $i' = j+1$,
or \emph{previous} if $j' + 1 = i$.

\begin{theorem}\label{th:above}
Suppose $\mmx{T}{T'}$.  
Then $\mcp{\bt{n}}{T'}$ is a tree partition of $T$ and $T'$.
\end{theorem}

\begin{proof}
Let $\mathcal{P} = \mcp{T}{T'}$, $\mathcal{Q} = \mcp{\bt{n}}{T'}$,
and $Q \in \mathcal{Q}$.
We show that $\mathcal{Q}$ is an enlargement of $\mathcal{P}$,
which implies $\even{T}{Q} = \even{T'}{Q}$,
and that $\trept{T}{Q} \subtr T$.

To start, we characterize how $\mathcal{Q}$ splits $T'$.
Let $S' = \trept{T'}{Q}$.
Since $\mmx{\bt{n}}{T'}$, then $S' \subtr T'$ consists 
of some odd siblings or an even parent with some odd children. 
We claim that an odd edge is in the same induced subtree as 
all its sibling along with its even parent (if it has one).

Let $1 \in Q$.  
Then $S' \fiso \tp{|Q|}$.
Consider orphan $(i,j) \in T' \setminus S'$ with least $i > 1$. 
But then its previous sibling $(k, i-1) \in S'$ which contradicts
alignment of $\mathcal{Q}$ since $(i-1, i) \in \bt{n}$.
Hence $S'$ consists of all orphans in $T'$.
By a similar argument, if $1 \notin Q$, then $S' \biso \bt{|Q|}$ 
consists of an even parent and all its odd children.

Suppose $(i,j) \in T'$ is an even edge with $j \in P \cap Q$
for $P \in \mathcal{P}$. 
But then $\trept{T'}{P} \biso \bt{|P|}$
since $\mmx{T}{T'}$ by assumption.
Hence $(i,j)$ is the parent in $\trept{T'}{P} \subseteq S'$,
so $P \subseteq Q$.
If $P = Q$, then 
$\trept{T}{Q} \subtr T$ also, and 
$\even{T}{Q} = \even{T'}{Q}$.

Otherwise, let $(i',j') \in S' \setminus \trept{T'}{P}$
with least $i' > i$, and consider $P' \in \mathcal{P}$ with $i' \in P'$.
Since $(i',j')$ is an odd child of $(i,j) \in T'$, then
$\trept{T'}{P'} \fiso \tp{|P'|}$.
Thus, $P' \subset Q$.  
Moreover, $\trept{T'}{P \cup P'} \subtr T'$ 
and $\biso \bt{|P| + |P'|}$ by construction.

We claim that $\trept{T}{P \cup P'} \subtr T$ also.
Note $\min{P'} = i'$ odd.  
Let $\max{P'} = j''$ for $1 < i < i' < j' \leq j'' < j < 2n$.
Since $\trept{T}{P'} \fiso \bt{|P'|}$, then $(i', j'') \in T$ 
is the odd parent.
By choice of $i'$, there exists $(i'-1, k) \in \trept{T}{P} \biso \tp{|P|}$ 
with $j'' < k \leq j$.
Hence, $(i', j'')$ is the first child of $(i-1, k)$, 
and $\trept{T}{P'}$ is connected to $\trept{T}{P}$.

Inductively, $Q$ is the union of $P \in \mathcal{P}$.
Since $\even{T}{P} = \even{T'}{P}$, then $\trept{T}{Q}$ is aligned
with $\trept{T'}{Q}$.
Connectivity of $\trept{T}{Q}$ follows by building the enlargement
in order of the missing odd children of $(i,j) \in T'$.
Such a child belongs to $\trept{T'}{P} \fiso \tp{|P|}$ for 
odd $P \in \mathcal{P}$.
But then $\trept{T}{P} \fiso \bt{|P|}$.
By choice of $P$,
the odd parent in each additional $\trept{T}{P}$ 
must be the first child of some even child,
or the next sibling of an odd parent, 
already in the growing induced subtree.
The case when $1 \in Q$ proceeds along similar lines beginning with $P \ni 1$.
\end{proof}

\begin{corollary}\label{th:upper}
Let $S \in \Tn$.
Then $\mmx{T}{S}$ and $\mmx{T'}{S}$ if and only if
there exists a tree partition $\mathcal{P}$ of $T$, $T'$, and $S$
such that,
for $P \in \mathcal{P}$,
$\trept{S}{P} \fiso \tp{|P|}$ for $P$ odd
and $\biso \bt{|P|}$ otherwise. 
\end{corollary}

\begin{proof}
Suppose $\mmx{T, T'}{S}$.
Let $\mathcal{P} = \mcp{\bt{n}}{S}$.
Then by the proof of Theorem~\ref{th:above}, $\mathcal{P}$ splits $S$
as desired.
Also $\trept{T}{P} \subtr T$, $\trept{T'}{P} \subtr T'$, and
$\even{T}{P} = \even{S}{P} = \even{T'}{P'}$.
For the converse,
it suffices to observe that
pairing exchanges on distinct aligned subtrees are independent.
\end{proof}

\noindent
Note that $\mcp{\bt{n}}{S}$ is an enlargement of both 
$\mcp{\bt{n}}{T}$ and $\mcp{\bt{n}}{T'}$.  
However, it is not necessarily the least enlargement in $\polat$.
For example, consider again
$T = \{(1,8), (2,7), (3,6), (4,5)\}$ and
$T' = \{(1,4), (2,3), (5, 8), (6,7)\}$.
Then $\mcp{\bt{n}}{T} = \{\{1\}, \{3,7\}, \{5\}\}$ whereas
$\mcp{\bt{n}}{T'} = \{\{1, 5\}, \{3\}, \{7\}\}$.
Their only forward V-path is through $\tp{n}$.

\begin{theorem}\label{th:lub}
There is a unique $S \in \Tn$ with $c(S)$ minimal 
such that $\mmx{T}{S}$ and $\mmx{T'}{S}$.
\end{theorem}

\begin{proof}
Let $\mathcal{P} = \mcp{T}{T'}$.
For $P \in \mathcal{P}$ with $|P| = p$,
let $P = \{i_1,\ldots, i_p\}$ and $\even{T}{P} = \{j_1, \ldots, j_p\}$ 
in increasing order.
If $P$ odd, then $i_1 < j_1 < i_2 < \ldots i_p < j_p$.
Otherwise, $j_1 < \ldots < i_p$.
Define $\lambda(P) = \{(i_k, j_k) \mid 1 \leq k \leq p\}$ for $P$ odd,
else $\{(j_1, i_p)\} \cup \{(i_k, j_{k+1}) \mid 1 \leq k < p \}$.
Let $S = \bigcup_{P \in \mathcal{P}} \lambda(P)$.

We claim $S \in \Tn$.
As constructed, each index from $[1,2n]$
appears in exactly one ordered pair, 
and $\lambda(P)$ contains no crossing.
Suppose there are $(i,j), (i',j') \in S$ with
$1 \leq i < i' < j < j' \leq 2n$
for distinct $P, P' \in \mathcal{P}$ with
$i, j \in P \cup \even{T}{P}$, $i', j' \in P' \cup \even{T}{P'}$.

Let $J = [i+1,j-1]$.
Consider $(k,l) \in \trept{T}{P'}$.
Suppose either $k \in J$, $l \in [j+1, 2n]$ or 
$k \in [1, i-1]$, $l \in J$.
However, such an edge obstructs the edge in $\trept{T}{P}$ with 
index $i$ from the one with $j$.
Hence an edge from $\trept{T}{P'}$ has both or neither indices
in $J$ which implies that $\trept{T}{P'\cap J} \subtr T$.
The same reasoning holds for $T'$, contradicting 
minimality of $P'$.

By construction, $\mathcal{P}$ is a tree partition of $S$ as well as 
$T$ and $T'$.
Moreover, $\trept{S}{P} \fiso \tp{p}$ for $P$ odd, and 
$\biso \bt{p}$ otherwise. 
Hence $\mmx{T, T'}{S}$.
Furthermore, $S$ is the only tree 
which meets the isomorphism requirements in Corollary~\ref{th:upper} 
using $\mathcal{P}$ as the tree partition.

Let $k$ be the number of even $P \in \mathcal{P}$.
Then $c(S) = \sum_{P\ \text{odd}} p + \sum_{P\ \text{even}} (p-1) = n - k$.
We claim this is least possible.

Suppose $\mmx{T, T'}{S'}$ for $S' \neq S$.
Let $\mathcal{Q}$ be a tree partition satisfying Corollary~\ref{th:upper}
for $S'$.
Then $\mathcal{Q}$ must be a strict enlargement of $\mathcal{P}$ 
in $(O_n, \cap)$.
Also $c(S') = n - k'$ for $\mathcal{Q}$ with $k'$ even parts.  
Let $Q = P \cup P'$ for $Q \in \mathcal{Q}$, $P, P' \in \mathcal{P}$.
If $P$ and $P'$ are both odd, then $\trept{S'}{Q} = \lambda(P) \cup \lambda(P)$.
Hence, by choice of $S'$, $k' < k$.
\end{proof}

Exchanging backward moves for forward, and the roles of the star trees,
we have the following dual versions of these results.

\begin{corollary}\label{th:below}
Suppose $\mmx{T}{T'}$.  
Then $\mcp{T}{\tp{n}}$ is a tree partition of $T$ and $T'$.
\end{corollary}

\begin{proof}
Subtrees in $T$ induced by $\mcp{T}{\tp{n}}$ consist 
of an odd parent and all its even children.
A similar argument to Theorem~\ref{th:above} shows that 
$\mcp{T}{\tp{n}}$ is an enlargement of $\mcp{T}{T'}$, 
and that the corresponding induced subsets of $T'$ are subtrees 
aligned with those in $T$.
\end{proof}

\begin{corollary}\label{th:lower}
Let $S \in \Tn$.
Then $\mmx{S}{T}$ and $\mmx{S}{T'}$ if and only if
there exists a tree partition $\mathcal{P}$ of $T$, $T'$, and $S$
such that,
for $P \in \mathcal{P}$,
$\trept{S}{P} \fiso \bt{|P|}$ for $P$ odd
and $\biso \tp{|P|}$ otherwise.
\end{corollary}

\begin{corollary}\label{th:glb}
There is a unique $S \in \Tn$ with $c(S)$ maximal 
such that $\mmx{S}{T}$ and $\mmx{S}{T'}$.
\end{corollary}

\begin{proof}
Let $\mathcal{P} = \mcp{T}{T'}$ and define $\gamma(P) = 
\{(i_1, j_p)\} \cup \{(j_k, i_{k+1}) \mid 1 \leq k < p \}$ for odd 
$P \in \mathcal{P}$,
and $\{(j_k, i_k) \mid 1 \leq k \leq p\}$ otherwise.
A similar argument to Theorem~\ref{th:lub} 
for $S = \bigcup_{P \in \mathcal{P}} \gamma(P)$ holds 
with $c(S) = \sum_{P\ \text{odd}} 1 + \sum_{P\ \text{even}} 0 = k$
where $k$ is now the number of odd $P$.
\end{proof}

Let $\lub{T}{T'}$ denote the tree from Theorem~\ref{th:lub}
and $\glb{T}{T'}$ the one from Corollary~\ref{th:glb}.
Then these are the ``apex'' of 
the forward and backward V-paths with the lowest branching barrier.
If the apex of a V-path is $S$, then its branching skew is 
$\abs{c(S) - \frac{n+1}{2}}$
and length is $\abs{c(T) + c(T') - 2 \cdot c(S)}$.
Hence, the minimal skew one is a function of the number of even
and odd parts in $\mcp{T}{T'}$, and so is the length.
At least one of the V-paths through $\lub{T}{T'}$ or $\glb{T}{T'}$
has length at most $n-1$, 
although the other orientation (i.e.\ forward/backward) could be longer.

\section{Connection with noncrossing partitions}\label{sec:order}

The relation $\mmx{T}{T'}$ is a partial order on \Tn\ 
with \Gn\ as its Hasse diagram and $c(T)$ as a rank function.
In other words,
$T'$ covers $T$ if $T' = \mu_{E}(T)$ where $E$ is
either an odd parent and even child or two even siblings,
i.e.\ the pairing exchange is a forward move and $c(T') = c(T) + 1$.
When viewed as a poset,
$\lub{T}{T'}$ and $\glb{T}{T'}$ are the least upper bound
and greatest lower bound, respectively.
It is worth noting the symmetry in their construction.

We show that this partial order is isomorphic to
the well-known lattice of noncrossing partitions, \NCn.
A partition of $[1,n]$ is \emph{noncrossing} if there does not exists
$1 \leq a < b < c < d \leq n$ such that $a,c$ are in one part and
$b,d$ in another.
Noncrossing partitions are still ordered under refinement.
The greatest lower bound remains the largest refinement.
However, the least upper bound is the smallest enlargement that
is also noncrossing.

\begin{theorem}
There is an order preserving bijection from \Tn\ under $\mmx{T}{T'}$
to \NCn.
\end{theorem}

\begin{proof}
Let $\nc{T}$ be the partition of $[1,n]$ obtained by
projecting $\mcp{\bt{n}}{T}$ down under $\theta: 2i-1 \rightarrow i$.
Let $P, P' \in \mcp{\bt{n}}{T}$.
Recall that $\trept{T}{P}$ consist
of an even parent and all its odd children, or all the orphan edges.

Suppose $\nc{T}$ has a crossing $1 \leq a < b < c < d \leq n$ 
with $a,c \in \theta(P)$, $b,d \in \theta(P')$, and $a, b$ least possible.
Let $i' = \min{P'}$, $j' = \max{P'}$.
Then $(i', j')$ is the even parent in $\trept{T}{P'} \biso \bt{|P'|}$
with $2a - 1 < i' = 2b - 2$ and $2d - 1 \leq j' \leq 2n-1$.
But then $(i',j')$ obstructs the edge with index $2c-1$ from 
the one with $2a-1$ in $\trept{T}{P} \subtr T$.
Contradiction.
Hence $\nc{T} \in \NCn$.

Suppose now $\mathcal{N} \in \NCn$.
Let $N \in \mathcal{N}$ and consider
$I_N = \{ 2i-1, 2i-2 \pmod{2n} \mid i \in N\}$.
Then $\min I_N$ is odd exactly when $1 \in N$.
Define the pairings $\lambda(N)$ on the ordered indices in $I_N$ 
as in the proof of Theorem~\ref{th:lub}, and let 
$T_{\mathcal{N}} = \bigcup_{N \in \mathcal{N}} \lambda(N)$.
We claim that $T_{\mathcal{N}} \in \Tn$.
If so, then $\mcp{\bt{n}}{T_{\mathcal{N}}}$ 
projects down to $\mathcal{N}$ by construction
since $\lambda(N) \fiso \tp{|N|}$ when $1 \in N$
and $\biso \bt{|N|}$ otherwise. 

Suppose there is $(i,j) \in \lambda(N)$, $(k,l) \in \lambda(N')$
with $1 \leq i < k < j < l \leq 2n$.
For $x \in \{i,j,k,l\}$, let $x'$ be $\frac{x+1}{2}$ if $x$ is odd,
and $\frac{x+2}{2}\pmod{n}$ otherwise.
Then $i', j' \in N$, $k',l' \in N'$ and either
$1 \leq i' < k' < j' < l' \leq n$ if $l \neq 2n$
or $1 = l' < i' < k' < j' \leq n$ otherwise.
Contradiction.
Thus $T_{\mathcal{N}} \in \Tn$ is the unique pre-image of $\mathcal{N}$.

Let $E$ be two unobstructed edges in $T$ and $T' = \mu_{E}(T)$.
Given how $\mcp{\bt{n}}{T}$ splits $T$,
this is a forward move if and only if distinct 
$P, P' \in \mcp{\bt{n}}{T}$ are involved.
As in the proof of Theorem~\ref{th:minmax}, 
$\mcp{\bt{n}}{T'} \setminus \mcp{\bt{n}}{T} = \{P \cup P'\}$ as a result.
But then $\mathcal{N}_{T'}$ covers $\mathcal{N}_{T}$ in \NCn,
and the bijection is order-preserving.
\end{proof}

An immediate consequence is that 
plane trees with $k$ odd edges are equinumerous with 
noncrossing partitions with $n - k + 1$ parts,
which are counted by the Narayana number
$N(n,k) = \frac{1}{n} \binom{n}{k} \binom{n}{k-1}$.  
This partition of \Tn\ differs from the common 
one according to $k$ leaves~\cite{dershowitz-zaks-80},
yielded by the classic bijections~\cite{dershowitz-zaks-86, prodinger-83}  
with \NCn.
However, a more recent enumerative result~\cite{liu-wang-li-14}
gives a bijection via vertices of odd distance from the root, 
and hence a Narayana decomposition with the same sets.

The correspondence has three related bijections:\
taking the minimal tree partition with $\tp{n}$
and/or using the even indices to partition $T \in \Tn$.
Moreover,
the connection between $\mcp{\bt{n}}{T}$ and $\mcp{T}{\tp{n}}$ 
yields insight into counting orbits in \NCn\ under Kreweras 
complementation~\cite{kcomp,kreweras-72}.

\section{Existence of bounded skew paths}\label{sec:bounded}

Although V-paths are well-characterized mathematically, 
their branching skew is biologically unfavorable.
Hence, 
we now show there are paths in \Gn, other than forward ones,
having the minimum possible branching skew.

Call a path from $T$ to $T'$ through $T_m$ \emph{bounded} if  
$c(T) - 1 \leq c(T_m) \leq c(T') + 1$
and \emph{tightly bounded} if $c(T) \leq c(T_m) \leq c(T')$.
The distinction accounts for $c(T') - c(T) \leq 1$.
In other words, such a path is bounded away from high branching 
skew trees to the extent possible given its start and end.

A \emph{planted} plane tree $T$ has a monovalent root vertex so $(1,2n) \in T$.
Call $T$ \emph{doubly planted} if $(2, 2n-1) \in T$ also.

\begin{lemma}\label{lem:dp}
If $1 < c(T) < n$, then there is a bounded path from $T$ to a doubly 
planted $T'$ with $c(T') = c(T)$.
\end{lemma}

\begin{proof}
Suppose $(1,2n) \in T$.
If $(2,2n-1) \notin T$, then a forward move on  
$(2,i), (j,2n-1) \in T$ yields a doubly planted $T''$ with $c(T'') = c(T) + 1$. 
There is a backward move on $T''$ to yield a suitable $T'$ unless 
$T'' \setminus \{(1,2n), (2,2n-1)\} \fiso \bt{n-2}$.
But then $T = \bt{n}$ and $c(T) = 1$.

Otherwise, a backward move on $(1,i), (j, 2n) \in T$ 
yields a planted $T''$ with $c(T'') = c(T) - 1$. 
If $T''$ is doubly planted, there is a suitable forward move unless
$T'' \setminus \{(1,2n), (2,2n-1)\} \fiso \tp{n-2}$.
But then $T = \tp{n}$ and $c(T) = n$.
Else, $2 < i < j < 2n-1$ and a forward move on the first and last
children of $(1,2n) \in  T''$ yields $T'$.
\end{proof}

\begin{lemma}\label{lem:bind}
If $c(T') - c(T) \leq 1$, then there is a bounded path from $T$ to $T'$.
Otherwise, there is a tightly bounded one.
\end{lemma}

\begin{proof}
Any forward path is tightly bounded.
Hence, consider $\notmmx{T}{T'}$ where 
$c(\bt{n}) = 1 < c(T) \leq c(T') < n = c(\tp{n})$.
The result holds for $n = 3, 4$ since 
there is either a bounded forward V-path through $\tp{n}$ or 
backward one through $\bt{n}$.

Suppose $c(T') - c(T) > 1$.  
Then there are $S, S' \in \Tn$ with $\mmx{T}{S}$, $\mmx{S'}{T'}$,
and $c(T) < c(S) = c(S') < c(T')$.
The existence of a bounded path from $S$ to $S'$ implies a tightly 
bounded one from $T$ to $T'$.
By the previous lemma, we may assume $S$ and $S'$ are doubly planted.
Keeping $(1,2n)$ and $(2,2n-1)$ fixed, inductively there is a 
bounded path from $\mathcal{G}_{n-2}$ connecting
$S$ and $S'$ in \Gn.
If $c(T') =  c(T) + 1$, then the same reasoning holds as for
$c(S) = c(S')$.
\end{proof}

While the skew of these paths is well-characterized, the length 
is not straightforward since the recursion has various dependencies.
A forward path not involving 
$\bt{n}$ or $\tp{n}$ has length at most $n-3$.
Let $g_n$ be the maximum length of a bounded path, or tightly bounded 
if possible, for $\notmmx{T}{T'}$ in \Gn.
Then $g_3 = 2$, $g_4 = 3$, and $g_{n} \leq (n-3) + 4 + g_{n-2}$.

\section{Existence of geodesics with bounded skew}\label{sec:geodesic}

Finally, we consider shortest paths, also called geodesics, in \Gn.
We show their length is determined by the 
number of simply aligned subsets, which ranges from $n$ (when $T = T'$)
down to $1$.

When there is only one, the two trees are called \emph{complementary},
consistent with lattice terminology.
The simplest example is $\tp{n}$ and $\bt{n}$, and
all other complementary pairs likewise~\cite{fpsac} have $c(T) + c(T') = n + 1$.
Moreover~\cite{fpsac},
the diameter of \Gn\ is $n - 1$, and is achieved by complementary trees.
Their V-paths necessarily pass through $\tp{n}$ and $\bt{n}$, so 
we are interested in alternative geodesics with bounded skew.

Note that removing the edge $(i,j)$ splits $T$ into two subtrees 
--- its descendents and the rest of $T$.
Denote the former as $\desc{T}{i}{j}$ and latter as 
$\notdesc{T}{i}{j}$.
One may be empty;
vacuously $\emptyset \subtr T$.

\begin{lemma}\label{lem:indstep}
Let $(i,j) \in T \setminus T'$.
Suppose either $\desc{T}{i}{j} \subtr T'$ or 
$\notdesc{T}{i}{j} \subtr T'$.
Then the edges in $T'$ with indices $i$ and $j$ are unobstructed.
\end{lemma}

\begin{proof}
Suppose $i$ pairs with $k$ and $j$ with $l$ in $T'$.
If $\notdesc{T}{i}{j} \subtr T'$, then $i < k < l < j$.
An obstructing edge must have one index in $[k+1,l-1]$ and the 
other in $[1, i-1] \cup [j+1, 2n]$.
But the latter is not possible, since any edges in $T'$ with 
an index $< i$ or $> j$ agrees with $T$ by assumption. 
When the ordering of indices is considered circularly, the case
$\desc{T}{i}{j} \subtr T'$ is symmetric.
\end{proof}

\begin{lemma}
If $T$ and $T'$ have $k$ simply aligned subsets,
then their geodesic has length $n - k$.
\end{lemma}

\begin{proof}
Construct a path
$T = T_0$, $T_1$, \ldots, $T_{n-k} = T'$ 
inductively by considering $(i,j) \in T' \setminus T_m$ with minimal 
$j - i \geq 1$.
Then $\desc{T'}{i}{j} \subtr T_m$. 
Let $E \subset T_m$ have $i, j \in \ind{E}$.
Hence $(i,j) \in T_{m+1} = \mu_E(T_m)$.
Since the number of common edges
increases monotonically to $n$,
the path is a geodesic with length $\sum_{S} (|S| - 1) = n - k$ where 
$S \subseteq T$ are the $k$ original simply aligned subsets with $T'$.
\end{proof}

Call an edge $(k,k+1)$ a \emph{stem}.  
Also define $(1,2n)$ to be one.
Let $e$ be a pairing between index $1 \leq i \leq 2n$ and $j = i+1 \pmod{2n}$.
If $e \notin T$, then Lemma~\ref{lem:indstep} applies.
Call $e$ a \emph{forward stem} if $i$ is odd, since the move 
is a forward one.
Dually, $e$ and the move are both \emph{backward} if $i$ is even.
Note that $(1,2n)$ is an odd edge but a backward stem.

For technical reasons, $T = \{(1,2)\}$ is considered
both a forward and backward stem.
Since every unrooted tree has at least two leaves,
$T \in \Tn$ has two or more stems when $n > 1$.

\begin{lemma}\label{lem:stem}
If $c(T) \geq \frac{n+1}{2}$, then $T$ has at least one forward stem.
Dually, $T$ has a backward stem when $c(T) \leq \frac{n+1}{2}$.
\end{lemma}

\begin{proof}
Let $n \geq 2$. 
Suppose $c(T) \geq \frac{n+1}{2}$, and consider 
$T' = \{(1,2k)\} \cup \desc{T}{1}{2k}$.
Assume $T'' = T \setminus T' \neq \emptyset$.
If $c(T) \geq \frac{n}{2} + 1$, then the result holds by induction on
either $c(T') \geq \frac{k+1}{2}$ or $c(T'') \geq \frac{n-k+1}{2}$.
If $c(T) = \frac{n+1}{2}$, then $n$ is odd and, we may assume, so is $k$.
Then $c(T') < \frac{k+1}{2}$ and $c(T'') < \frac{n-k+1}{2}$ implies
$c(T') \leq \frac{k-1}{2}$ and $c(T'') \leq \frac{n-k}{2}$, a contradiction.
When $k = n$, the result holds by induction on 
$\desc{T}{1}{2n} \biso T'' \in \mathcal{T}_{n-1}$ 
with $c(T'') = n - c(T) \leq \frac{n-1}{2}$.
The dual result follows from the mapping $i \rightarrow i + 1 \pmod{2n}$
on the half-edge indices,
which is a bijection on \Tn.
The image has $n - c(T) + 1$ odd edges, and the forward/backward 
orientation of stems reversed.
\end{proof}

\begin{lemma}\label{lem:zigzag}
If $T, T'$ are complementary with $c(T) = c(T')$, 
then they have a bounded geodesic.
\end{lemma}

\begin{proof}
Since $c(T) + c(T') = n+1$, 
consider odd $n \geq 3$.
By Lemma~\ref{lem:stem},
$T'$ has both a forward and backward stem.
The corresponding moves on 
$E \subset T$ and $F \subset \mu_{E}(T)$
yield $T'' = \mu_{F}(\mu_{E}(T))$ with $c(T'') = \frac{n+1}{2}$.
Then $\mcp{T''}{T'}$ has three parts, two of which are singletons.
Let $P$ be the non-singleton.
Then $\trept{T''}{P}$ and $\trept{T'}{P}$ are complementary 
with $\frac{n-1}{2}$ odd edges.
Inductively, their images in $\mathcal{G}_{n-2}$ have a bounded geodesic.
Keeping common edges fixed, so do $T$ and $T'$.
\end{proof}

\noindent
In other words, it is possible to ``zigzag'' between complementary 
$T$ and $T'$ when $c(T) = c(T')$.
Since the two moves can be made in either order,
when $c(T) < c(T')$, they can be sequenced 
not to exceed the original skew.

\begin{lemma}\label{lem:nonzigzag}
If $T, T'$ are complementary with $c(T) \neq c(T')$, 
they have a tightly bounded geodesic.
\end{lemma}

\begin{proof}
Suppose $n \geq 4$
and $c(T) < \frac{n+1}{2} < c(T')$.
There is a forward move on $E \subset T$ corresponding to $(i,i+1) \in T'$.
Let $S = \mu_{E}(T) \setminus \{(i,i+1)\}$
and $S' = T' \setminus \{(i,i+1)\}$ be the resulting complementary
subtrees.
If $c(T) = c(S) < \frac{n}{2} < c(S') = c(T') - 1$, 
the result holds inductively.
Else, $\mu_{E}(T) = \frac{n}{2} + 1 = c(T')$,
and Lemma~\ref{lem:zigzag} applies to $S$ and $S'$.
By applying the backward move first to $\mu_{E}(T)$,
the geodesic from $T$ to $T'$ will be tightly bounded.
\end{proof}

These results extend directly when there is a bijection between 
simply aligned subsets and parts of the minimal tree partition. 
Call simply aligned $S \subseteq T$, $S' \subseteq T'$ 
a \emph{block} if $S \subtr T$ and $S' \subtr T'$.
Say $T$ and $T'$ have a \emph{block decomposition} when the induced 
subtrees from $\mcp{T}{T'}$ are simply aligned,
i.e.\ complementary.
Call a pairing exchange a \emph{geodesic} move if it maintains 
a block decomposition while increasing the number of common edges. 

\begin{lemma}
Suppose $T$ and $T'$ have a block decomposition.
If $c(T') = c(T)$, then there is a bounded geodesic from $T$ to $T'$.
Otherwise, there is a tightly bounded one.
\end{lemma}

\begin{proof}
Let $S_i \subseteq T$, $S'_i \subseteq T'$ be the simply aligned pairs.
Since $S_i \subtr T$ and $S'_i \subtr T'$,
each pair can be treated independently.

Suppose $c(T) = c(T')$.
If $c(S_i) \neq c(S'_i)$, then Lemma~\ref{lem:nonzigzag} applies.
Since $\sum c(S_i) = \sum c(S'_i)$, alternate a geodesic forward move for $T$
on $S_i$ where $c(S_i) < c(S'_i)$ with a backward one on $S_j$ where 
$c(S_j) > c(S'_j)$ 
until Lemma~\ref{lem:zigzag} applies to all pairs.

If $c(T) = c(T') - 1$, again alternate moves 
until Lemma~\ref{lem:zigzag} applies 
to all pairs but $c(S_i) = c(S'_i) - 1$.
Then, as in the proof of Lemma~\ref{lem:nonzigzag}, the geodesic moves
on $S_i$ and the other pairs 
can be sequenced so that the path is tightly bounded.

Suppose $c(T') - c(T) > 1$.
Then either $c(S_i) + 2 \leq c(S'_i)$ or  
$c(S_i) + 1 \leq c(S'_i)$ and $c(S_j) + 1 \leq c(S'_j)$.
But then there is a geodesic forward move on $T$ and a backward one 
on $T'$ which decreases $c(T') - c(T)$ by 2.
Keeping common edges fixed, 
inductively applying any of these cases will not increase the skew beyond
the original bounds.
\end{proof}

\noindent
The case when $c(T') -  c(T) = 1$ differs from Lemma~\ref{lem:bind} 
because the moves for Lemma~\ref{lem:dp} are ordered, 
unlike Lemma~\ref{lem:zigzag}.
When $T$ and $T'$ do not have a block decomposition, 
moves on simply aligned subsets are not necessarily independent. 
Hence, the sequencing becomes more complicated,
and such bounds may not hold in general.

\section{Acknowledgments}

The authors thank Sinan Aksoy and Stephen Young 
for all their efforts in promoting applied combinatorics.
Thanks are also due to Sergey Fomin for making explicit the connection
with noncrossing partitions. 
This work was supported by the Burroughs Wellcome Fund (2005 CASI to CH),
National Science Foundation (DMS1815044 to CH), and 
National Institutes of Health (R01GM126554 to CH).


\begin{thebibliography}{10}

\bibitem{ldpmath}
Y.~Bakhtin and C.~Heitsch.
\newblock Large deviations for random trees.
\newblock {\em J Stat Phys}, 132(3):551--560, 2008.

\bibitem{ldpbio}
Y.~Bakhtin and C.~Heitsch.
\newblock Large deviations for random trees and the branching of {RNA}
  secondary structures.
\newblock {\em Bull. Math. Biol.}, 71(1):84--106, 2009.

\bibitem{regions}
F.~Barrera-Cruz, C.~Heitsch, and S.~Poznanovi\'c.
\newblock On the structure of {RNA} branching polytopes.
\newblock {\em SIAM J Appl Algebra Geometry}, 2(3):444--461, 2018.

\bibitem{bevilacqua-blose-08}
P.~C. Bevilacqua and J.~M. Blose.
\newblock Structures, kinetics, thermodynamics, and biological functions of
  {RNA} hairpins.
\newblock {\em Annu Rev Phys Chem}, 59:79--103, 2008.

\bibitem{dershowitz-zaks-80}
N.~Dershowitz and S.~Zaks.
\newblock Enumerations of ordered trees.
\newblock {\em Discrete Math.}, 31(1):9--28, 1980.

\bibitem{dershowitz-zaks-86}
N.~Dershowitz and S.~Zaks.
\newblock Ordered trees and noncrossing partitions.
\newblock {\em Discrete Math.}, 62(2):215--218, 1986.

\bibitem{dotu-etal-10}
I.~Dotu, W.~A. Lorenz, P.~V. Hentenryck, and P.~Clote.
\newblock Computing folding pathways between {RNA} secondary structures.
\newblock {\em Nucleic Acids Res}, 38(5):1711--22, 2010.

\bibitem{gan-pasquali-schlick-03}
H.~H. Gan, S.~Pasquali, and T.~Schlick.
\newblock Exploring the repertoire of {RNA} secondary motifs using graph
  theory; implications for {RNA} design.
\newblock {\em Nucleic Acids Res}, 31(11):2926--43, 2003.

\bibitem{kcomp}
C.~Heitsch.
\newblock Counting orbits under {K}reweras complementation.
\newblock In preparation.

\bibitem{dna8}
C.~Heitsch, A.~Condon, and H.~Hoos.
\newblock From {RNA} secondary structure to coding theory: A combinatorial
  approach.
\newblock In A.~O. M.~Hagiya, editor, {\em {DNA8}: Revised Papers from the 8th
  International Workshop on DNA Based Computers}, volume 2568 of {\em Lecture
  Notes in Computer Science}, pages 215--228, London, UK, 2003.
  Springer-Verlag.

\bibitem{insights_chapt}
C.~Heitsch and S.~Poznanovi\'{c}.
\newblock Combinatorial insights into {RNA} secondary structure.
\newblock In N.~Jonoska and M.~Saito, editors, {\em Discrete and topological
  models in molecular biology}, Nat. Comput. Ser., pages 145--166. Springer,
  Heidelberg, 2014.

\bibitem{fpsac}
C.~Heitsch and P.~Tetali.
\newblock Meander graphs.
\newblock In {\em 23rd {I}nternational {C}onference on {F}ormal {P}ower
  {S}eries and {A}lgebraic {C}ombinatorics ({FPSAC} 2011)}, Discrete Math.
  Theor. Comput. Sci. Proc., AO, pages 469--480. Assoc. Discrete Math. Theor.
  Comput. Sci., Nancy, 2011.

\bibitem{parametric}
V.~Hower and C.~Heitsch.
\newblock Parametric analysis of {RNA} branching configurations.
\newblock {\em Bull Math Biol}, 73(4):754--76, 2011.

\bibitem{huston-etal-21}
N.~C. Huston, H.~Wan, M.~S. Strine, R.~de~Cesaris Araujo~Tavares, C.~B. Wilen,
  and A.~M. Pyle.
\newblock Comprehensive in vivo secondary structure of the {SARS-CoV-2} genome
  reveals novel regulatory motifs and mechanisms.
\newblock {\em Mol Cell}, 81(3):584--598.e5, 2021.

\bibitem{kreweras-72}
G.~Kreweras.
\newblock Sur les partitions non crois\'ees d'un cycle.
\newblock {\em Discrete Math.}, 1(4):333--350, 1972.

\bibitem{li-zhang-12}
Y.~Li and S.~Zhang.
\newblock Predicting folding pathways between {RNA} conformational structures
  guided by {RNA} stacks.
\newblock {\em BMC Bioinformatics}, 13(Suppl 3):S5, 2012.

\bibitem{liu-wang-li-14}
C.~Liu, Z.~Wang, and B.~Li.
\newblock Bijections between bicoloured ordered trees and non-crossing
  partitions.
\newblock {\em Ars Combin.}, 117:155--162, 2014.

\bibitem{morgan-higgs-98}
S.~R. Morgan and P.~G. Higgs.
\newblock Barrier heights between groundstates in a model of {RNA} secondary
  structure.
\newblock {\em J Phys A (Math \& General)}, 31(14):3153--3170, 1998.

\bibitem{petrov-etal-13}
A.~S. Petrov, C.~R. Bernier, E.~Hershkovits, Y.~Xue, C.~C. Waterbury, C.~Hsiao,
  V.~G. Stepanov, E.~A. Gaucher, M.~A. Grover, S.~C. Harvey, N.~V. Hud, R.~M.
  Wartell, G.~E. Fox, and L.~D. Williams.
\newblock Secondary structure and domain architecture of the {23S} and {5S}
  {rRNAs}.
\newblock {\em Nucleic Acids Res}, 41(15):7522--7535, 2013.

\bibitem{robust}
S.~Poznanovi\'c, F.~Barrera-Cruz, A.~Kirkpatrick, M.~Ielusic, and C.~Heitsch.
\newblock The challenge of {RNA} branching prediction: a parametric analysis of
  multiloop initiation under thermodynamic optimization.
\newblock {\em J Struct Biol}, 210(1):107475, 2020.

\bibitem{bnb}
S.~Poznanovi\'c, C.~Wood, M.~Cloer, and C.~Heitsch.
\newblock Improving {RNA} branching predictions: Advances and limitations.
\newblock {\em Genes (Basel)}, 12(4):469, 2021.

\bibitem{prodinger-83}
H.~Prodinger.
\newblock A correspondence between ordered trees and noncrossing partitions.
\newblock {\em Discrete Math.}, 46(2):205--206, 1983.

\bibitem{quinn-etal-14}
J.~J. Quinn, I.~A. Ilik, K.~Qu, P.~Georgiev, C.~Chu, A.~Akhtar, and H.~Y.
  Chang.
\newblock Revealing long noncoding {RNA} architecture and functions using
  domain-specific chromatin isolation by {RNA} purification.
\newblock {\em Nat Biotechnol}, 32(9):933--940, 2014.

\bibitem{schmitt-waterman-94}
W.~R. Schmitt and M.~S. Waterman.
\newblock Linear trees and {RNA} secondary structure.
\newblock {\em Discrete Appl. Math.}, 51(3):317--323, 1994.

\bibitem{stanley-99}
R.~P. Stanley.
\newblock {\em Enumerative combinatorics. {V}ol. 2}, volume~62 of {\em
  Cambridge Studies in Advanced Mathematics}.
\newblock Cambridge University Press, Cambridge, 1999.

\bibitem{takizawa-etal-20}
H.~Takizawa, J.~Iwakiri, G.~Terai, and K.~Asai.
\newblock Finding the direct optimal {RNA} barrier energy and improving
  pathways with an arbitrary energy model.
\newblock {\em Bioinformatics}, 36(Suppl\_1):i227--i235, 2020.

\bibitem{turner-mathews-10}
D.~H. Turner and D.~H. Mathews.
\newblock {NNDB}: the nearest neighbor parameter database for predicting
  stability of nucleic acid secondary structure.
\newblock {\em Nucleic Acids Res}, 38:D280--2, 2010.

\bibitem{sulc-20}
P.~\v{S}ulc.
\newblock The multiscale future of {RNA} modeling.
\newblock {\em Biophys J}, 119(7):1270--1272, 2020.

\bibitem{xu-chen-12}
X.~Xu and S.-J. Chen.
\newblock Kinetic mechanism of conformational switch between bistable {RNA}
  hairpins.
\newblock {\em J Am Chem Soc}, 134(30):12499--12507, 2012.

\end{thebibliography}

\end{document}